\documentstyle[epsfig]{mn}
\title[Observations of HDF with ISO IV ]
{Observations of the Hubble Deep Field with the Infrared  Space
Observatory IV: Association of sources with Hubble Deep Field galaxies}

\author[R.G. Mann et al.]
{R.G. Mann$^1$, S.J. Oliver$^1$, S.B.G. Serjeant$^1$, M. Rowan-Robinson$^1$
\vspace*{0.2cm}\\{\LARGE
A. Baker$^3$, N. Eaton$^1$, A. Efstathiou$^1$, P. Goldschmidt$^1$, 
}\vspace*{0.2cm}\\{\LARGE
B. Mobasher$^1$, T. Sumner$^1$, L. Danese$^2$, D. Elbaz$^3$,
}\vspace*{0.2cm}\\{\LARGE
A. Franceschini$^4$, E. Egami$^5$, M. Kontizas$^6$, A. Lawrence$^7$,
}\vspace*{0.2cm}\\{\LARGE
R. McMahon$^8$, H.U. Norgaard-Nielsen$^9$, I. Perez-Fournon$^{10}$, 
}\vspace*{0.2cm}\\{\LARGE
I. Gonzalez-Serrano$^{11}$
}\\  
$^1$Astrophysics Group, Blackett Laboratory, Imperial College, 
Prince Consort Road, London SW7 2AZ\\
$^2$SISSA, Via Beirut 2-4,Trieste, Italy\\
$^3$Service d'Astrophysique, Saclay, 91191 Gif-sur-Yvette, Cedex, France\\
$^4$Osservatorio Astronomico di Padova, Vicolo dell'Osservatorio 5,
I-35 122, Padova, Italy\\
$^5$Max-Planck-Institut f\"ur Extraterrestrische Physik,
Giessenbachstrasse, D-8046, Garching bei Munchen, Germany\\
$^6$Astronomical Institute, National Observatory of Athens, P.O. Box
200048, GR-118 10,  Athens, Greece\\
$^7$Institute for Astronomy, University of Edinburgh, Blackford Hill,
Edinburgh, EH9 3HJ\\
$^8$Institute of Astronomy, The Observatories, Madingley
Road,Cambridge, CB3 0HA\\
$^9$Danish Space Research Institute, Gl. Lundtoftevej 7, DK-2800
Lyngby, Copenhagen, Denmark\\
$^{10}$Instituto Astronomico de Canarias, Via Lactea, E-38200 La
Laguna, Tenerife, Canary Islands, Spain\\
$^{11}$Instituto de Fisica de Cantabria, Santander, Spain\\
}

\newcommand{\beq}{\begin{equation}}
\newcommand{\eeq}{\end{equation}}
\newcommand{\bfi}{\begin{figure}}
\newcommand{\efi}{\end{figure}}
\newcommand{\bit}{\begin{itemize}}
\newcommand{\eit}{\end{itemize}}

\newcommand{\myref}[1]{\noindent \hangindent=0.5in \hangafter=1 #1 \par}
\newcommand{\gs}{\mathrel{\raise1.16pt\hbox{$>$}\kern-7.0pt\lower3.06pt\hbox
{{$\scriptstyle \sim $}}}}
\newcommand{\ls}{\mathrel{\raise1.16pt\hbox{$<$}\kern-7.0pt\lower3.06pt\hbox
{{$\scriptstyle \sim $}}}}
\renewcommand{\baselinestretch}{1.0}

\begin{document}

\maketitle
\renewcommand{\baselinestretch}{1.0}
\begin{abstract}
We discuss the identification of sources detected by ISO at 6.7 and 15$\mu$m in
the Hubble Deep Field (HDF) region. We conservatively associate ISO sources with 
objects in existing optical and near-infrared HDF catalogues using the likelihood ratio
method, confirming these results (and, in one case, clarifying
them) with independent visual searches.
We find fifteen ISO sources to be reliably associated with bright \mbox{[$I_{814}(AB) < 23$]} 
galaxies in the HDF, and one with an \mbox{$I_{814}(AB)=19.9$} star,
while a further eleven are associated with objects in the Hubble
Flanking Fields (ten galaxies and one star). Amongst
optically bright HDF galaxies, ISO tends to detect luminous,
star-forming galaxies at fairly
high redshift and with disturbed morphologies, in preference to nearby ellipticals.
\end{abstract} 

\section{INTRODUCTION}

This series of papers describes a set of very deep  mid-infrared observations,
obtained using the ISOCAM (Cesarsky et al. 1996) 
instrument on the Infared Space Observatory (ISO, Kessler et al. 1996) and
centred on the Hubble Deep Field (HDF, Williams et al. 1996) region. In 
Paper I (Serjeant et al. 1997) we discussed the reduction of the ISOCAM data
and presented the  resultant maps of the HDF region at 6.7$\mu$m and
15$\mu$m. Paper II (Goldschmidt et al. 1997) 
described the methods we used for detecting sources in these maps,
while Paper III (Oliver et al. 1997) compared source counts derived from
these detections with model predictions. 
The two principal goals of this paper are to confirm the ISO-HDF
sources, through associating them with objects in  existing HDF
galaxy catalogues, and to study the properties of those associated
galaxies, contrasting them with those of bright HDF galaxies
not detected by ISO.
The spectral energy distributions resulting from the association
procedure will be discussed in Paper V (Rowan-Robinson et al. 1997),
together with their implications for the star formation history of the
Universe.

The plan of this paper is as follows. In Section 2, we briefly review
the basic problems we face in associating the ISO-HDF sources with
galaxies in existing HDF catalogues. The
likelihood ratio method for source 
identification is described in Section 3, where we present the results
of applying it to our ISO-HDF sources and discuss the reliability of
the associations we made using it. In Section 4 we discuss the properties
of the galaxies associated with our ISO sources, as well as those
prominent optical HDF galaxies we did not detect, and present
the conclusions that we draw from the work described in this
paper.

\section{ISO SOURCE DETECTIONS IN THE HUBBLE DEEP FIELD}

A total of 27 sources (7 in the complete sample, and 20 in the
supplementary sample) were detected in the 6.7 $\mu$m map together with 22
sources (19 complete, 3 supplementary) at 15 $\mu$m: the positions 
and fluxes of these objects are tabulated in Paper II. 

There are several basic problems which complicate 
the association of ISO-HDF sources with HDF galaxies in existing
optical and near-infrared catalogues. The most obvious of these is the
poor match between
the  resolution of ISOCAM and the high source density of
galaxies in the HDF: the radius of the Airy disk is 2.8 arcsec at 
6.7$\mu$m and 6.0 arcsec at 15$\mu$m, while there are several hundred
galaxies per square arcminute detected 
to $I_{814}\sim 29$, so we expect several galaxies per
randomly placed Airy disk at both  6.7 $\mu$m and 15 $\mu$m. 
({\em N.B.\/} Unless otherwise stated, the magnitudes used in this
paper are total magnitudes in the AB system, as measured by Williams
et al. 1996, and we follow them in denoting their four optical bands
as $U_{300}$, $B_{450}$, $V_{606}$ and $I_{814}$ to avoid confusion
with the standard Johnson photometric bands.)
Thus, not only is there a high likelihood of chance
associations with optical galaxies, but any given ISOCAM beam may be
integrating over more than one source, and significant 
flux may be contributed by more than one galaxy: this latter is
exacerbated both because many luminous mid-infrared sources are
likely to be interacting or merging galaxy systems (e.g. Lawrence
et al. 1989), and by the fact that our ISO maps appear to be at, or close
to, the confusion limit in both bands (Paper I).
An additional positional uncertainty results from the possibility
of field distortions in the original ISO data (Paper I). We
make some allowance for this in our source association procedure (see 
Section 3), and feel that it is unlikely to affect our results
significantly. 
Another issue is the band in which to make the associations, since,
clearly, the wide range of colours exhibited by HDF galaxies could
mean that different galaxies would be associated with a particular
ISO source in different bands, and an unfortunate choice of band 
could bias the associations made: for example, we might expect the true
counterparts of our ISO galaxies to be dusty, so using too blue a
band might lead us to make the wrong associations. With that in mind we have performed
the likelihood ratio procedure of Section 3 on the $I_{814}$ images
of Williams et al. (1996), which is both the reddest and deepest band,
as well as the only one
available in the Hubble Flanking Fields (HFF).
Finally, we have relatively few sources (only 19 of 
the 6.7$\mu$m sources and 5 of those at 15$\mu$m have Airy disks which
fall within
the HDF), restricting our ability to use statistical techniques which rely on
the determination of properties of the source population from the data
themselves.

\begin{figure}
\epsfig{file=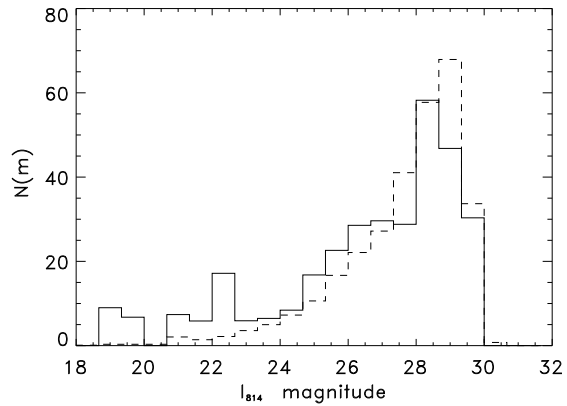,angle=0,width=8cm}
\caption{$I_{814}$ band magnitude histograms for galaxies near ISO-HDF
6.7$\mu$m source positions, and for the whole Williams et al. (1996)
catalogue. The solid lines show the magnitude distribution of galaxies
lying within twice the Airy disk radius of the 19 6.7$\mu$m sources
whose Airy disks fall within
in the HDF, while the dashed line traces the magnitude distribution for the full HDF
catalogue, normalised to the total area enclosed by the 19 Airy
disks. There is a clear excess of bright galaxies surrounding the
ISO-HDF sources.}
\end{figure}

\begin{figure}
\epsfig{file=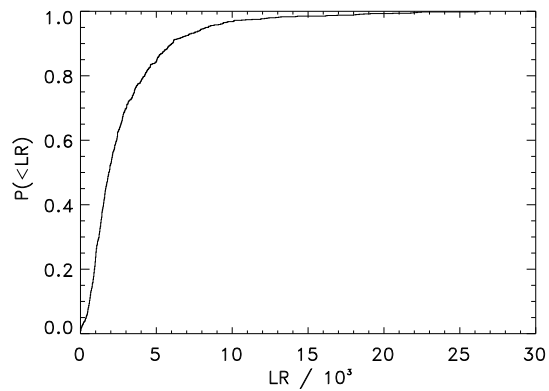,angle=0,width=8cm}
\caption{The cumulative probability distribution for the likelihood
ratio of the likeliest optical counterpart to a fictitious 6.7$\mu$m
source placed at random in the HDF.}
\end{figure}

As a first step towards our goals, we compare the $I_{814}$ 
band magnitude 
distribution of galaxies near our ISO-HDF source positions with that
of the complete
optical HDF galaxy catalogue of Williams et al. (1996). In Fig. 1
we plot the $I_{814}$ band magnitude distribution of those HDF 
galaxies within
twice the Airy disk radius of the 19 6.7$\mu$m ISO-HDF source
positions lying 
within the HDF, together with that for the full Williams et al. (1996) optical
galaxy catalogue. The histogram for the ISO-HDF neighbours is noisy,
both because of the small number of sources, and because a number of
them lie close to the edge of the HDF, and the correction made
for the fraction of the search region outside the HDF can give large
weights to those HDF galaxies inside it. It is clear
from Fig. 1 that there is an excess of bright ($I_{814} < 23$)
galaxies surrounding our ISO-HDF source positions: a two-sided 
Kolmogorov-Smirnov test yields a probability $P=1.6 \times 10^{-3}$
that the two magnitude distributions are drawn from the same
population (falling to $P=5.4 \times 10^{-4}$ when only the
six sources from the complete 6.7$\mu$m sample are considered), and
the five 15$\mu$m sources in the HDF yield similar results. This 
strongly suggests that the sources in our ISO-HDF samples are 
associated with bright galaxies in the HDF.

\begin{figure*}
\epsfig{file=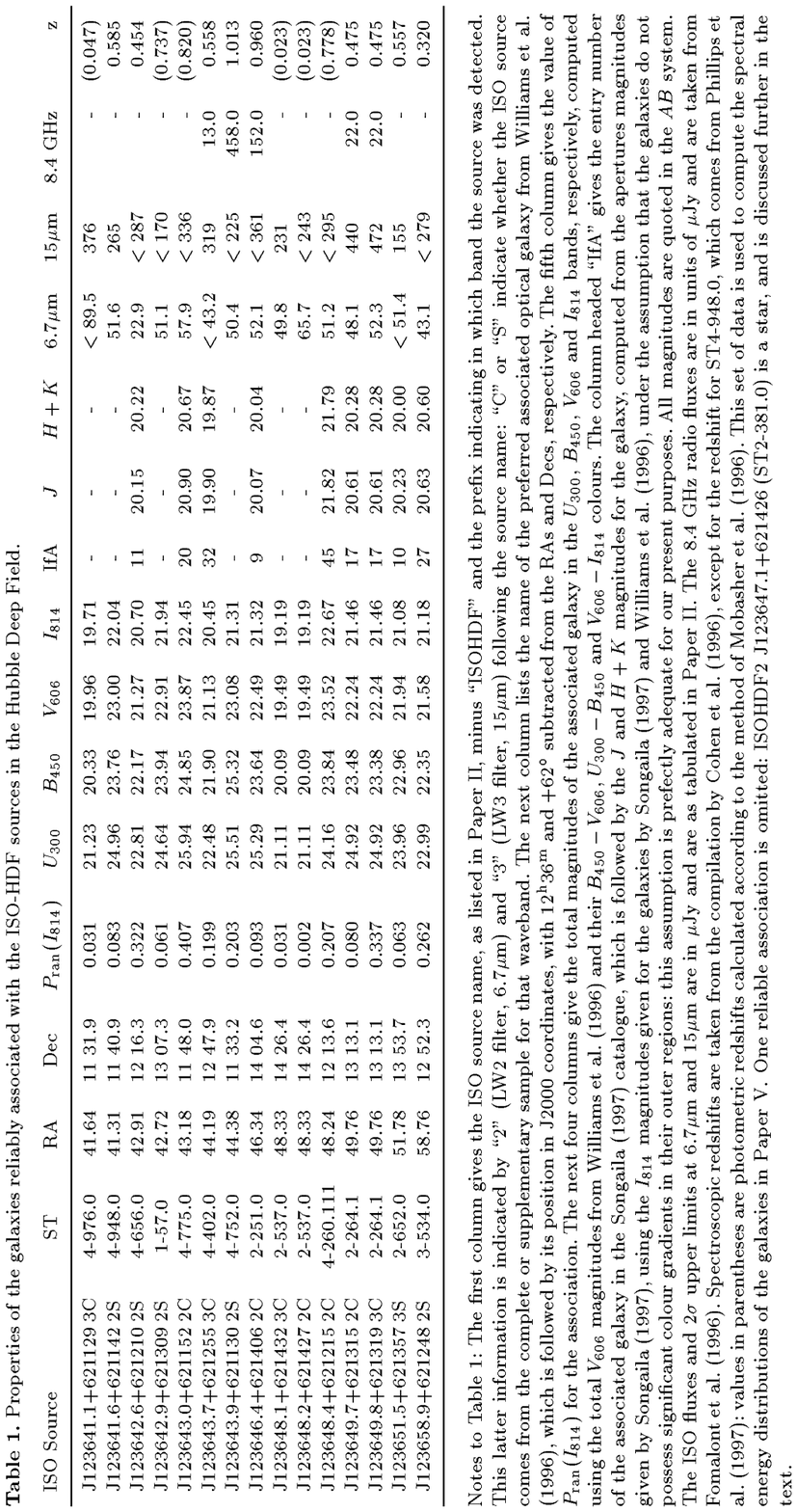,angle=180,width=12.0cm}
\end{figure*}

\begin{figure*}
\vspace{5cm}
\caption{Positions of ISO-HDF sources and their associations, from
a $I_{814}$ band mosaic of the HDF. Dashed circles show the Airy disks of
the ISO-HDF sources, squares mark the positions of the associated
optical galaxies: a second dashed circle indicates a second source
from the same sample nearby. Where the ISO source position falls
within the HDF, the plots are centred on those positions: in three
cases the source falls just outside the HDF and the Airy disk is 
displaced from the centre of the plot. The title to each plot gives the name
of the ISO source and the Williams et al. (1996) name for the optical
galaxy associated with it.}
\end{figure*}

\section{ASSOCIATING ISO-HDF SOURCES USING THE LIKELIHOOD RATIO METHOD}

\subsection{The likelihood ratio method}
	
The likelihood ratio method is one of the most commonly used
techniques for associating {\em sources\/} in one catalogue with
{\em objects\/} in another: it is described in detail by 
Sutherland \& Saunders (1992), so we present only a brief review here.
The likelihood ratio, $LR$, is defined to be the ratio of the probability,
$p_{\rm true}$,
of finding the true counterpart to the source at the position of the 
object and with its flux, to the probability, $p_{\rm chance}$, of
finding a chance
object at that position and with that flux, given the errors in the 
source and object positions. 
Consider an object with positional offsets
$(x,y)$ from the estimated source position and with flux $f$. The
probability that the true counterpart lies in an infinitesimal region
of area ${\rm d}x{\rm d}y$ about that position and has a flux in an
infinitesimal interval of size ${\rm d}f$ about that flux, is given by
\begin{equation}
p_{\rm true} \; = \; q(f) \: e(x,y) \: {\rm d}f \: {\rm d}x \: {\rm d}y,
\end{equation}
where $e(x,y)$ is the joint probability distribution function for $x$
and $y$, normalised so that \mbox{$\int e(x,y) {\rm d}x {\rm
d}y=1$} and $q(f)$ is the probability distribution function for an
ensemble of sources, measured in the passband in which the object
catalogue is defined. If $n(f)$ is the local surface density of
objects per unit flux, then \mbox{$p_{\rm chance}=n(f){\rm d}f {\rm d}x
{\rm dy}$} and $LR$ is given by 
\begin{equation}
LR(f,x,y) \; = \; \frac {q(f) \: e(x,y)}{n(f)}.
\label{LR_final}
\end{equation}
To implement this method to associate ISO-HDF sources with objects in
the STScI HDF optical catalogue (Williams et al., 1996), we make the following
assumptions concerning the quantities present in equation
(\ref{LR_final}). We neglect the uncertainties in the optical
positions, setting $e(x,y)$ equal to
a Gaussian distribution, with a $\sigma$ value equal to the quadrature
sum of the radius of the Airy disk for the ISO sources (2.8 arcsec at
7$\mu$m and  6.0 arcsec at 15$\mu$m) and an estimated positional
error of equal size: this is a crude estimate of the true
positional uncertainties, designed, in part, to take account of the
possibility of ISOCAM field distortions, but the
associations made are
insensitive to variation of this figure within reasonable bounds. The
form of $n(f)$, which is the magnitude distribution of the galaxies in
the $I_{814}$ band, is
readily computed, but the choice of $q(f)$ is more problematic. As
discussed by Sutherland \& Saunders (1992), the two conventional
approaches would involve assuming some model for the
magnitude distribution of the true optical counterparts of our ISO-HDF
sources, or estimating $q(f)$ from the data, which, essentially, means
taking the difference between the pair of histograms shown in Fig. 1. The
latter method would, clearly, yield an unsatisfactorily noisy $q(f)$
(particularly for the 15$\mu$m sources), while sufficiently little is known
about the mid-infrared properties of the galaxies we are likely to
detect with ISO in the HDF that adopting a model based, say, on IRAS data, 
could seriously bias our results. We choose, instead, to take $q(f)$ equal
to a constant, independent of magnitude: uncertainty as to the exact
reliability of our ISO samples means that the value this constant should take remains
unclear, so that the likelihood ratios we compute are left
unnormalised, proportional to those defined by equation (\ref{LR_final}).
This uniform $q(f)$ is, no doubt, incorrect, since, on the evidence of
Fig. 1 (as well as {\em a priori\/} prejudice) the galaxies we detect are
likely to be amongst the brightest in the HDF. Its effect, given that point,
is to make us more likely to identify our ISO sources with
fainter galaxies than we should be. In 
fact, as we shall see, our associations are with bright galaxies,
which suggests that our taking a uniform $q(f)$ has not biased
our results.

\subsection{Results}

Deciding what level of likelihood ratio we consider to correspond to
a reliable association is a somewhat subjective matter. Independent of the
application of the likelihood ratio technique, two of us (RGM and
MRR) made independent associations by eye, using $I_{814}$ images of
the HDF fields, with the Airy disks of the ISO-HDF sources superimposed
upon them. In making these associations we are likely to have been
making a subconscious balance between the brightness of a given galaxy and
its distance from the source position, qualitatively similar
to the likelihood ratio method, where $LR \propto
e(x,y)/n(f)$. Nevertheless, the level of agreement obtained, both 
between the two observers and with the results of the likelihood ratio
analysis, is surprising: for all but one source the two observers
agreed on the most likely association, and the sources which they felt
confident in having associated reliably were in a one-to-one
correspondence with the sources yielding the highest likelihood
ratios. 

We can estimate the level of reliability of these associations by
considering the likelihood ratios for association with random points
in the HDF. In Fig. 2 we show the cumulative probability
distribution for the likelihood ratio of the most likely HDF galaxy for being
the optical counterpart of a fictitious 6.7$\mu$m source at a random position
in the HDF, computed using the same forms for $q(f)$ and $n(f)$ as
for the ISO-HDF sources and 10000 random positions in the HDF. 
From Fig. 2 we can compute, as a function of $LR$, the quantity 
$P_{\rm ran}(I_{814})$, which is the probability that a
fictitious source, placed at random in the HDF, would have a likeliest
association in the $I_{814}$ catalogue of Williams et al. (1996) with
a likelihood ratio at least as high as a given value
of $LR$. The $P_{\rm ran}(I_{814})$ values for the best optical candidates for
the 6.7$\mu$m sources marking the boundary above which the observers
conservatively considered their associations to be reliable was about
$P_{\rm ran}=0.35$. We chose this conservative level to mark the
break between those associations we consider to be reliable and those
we do not.

In Table 1 we list the properties of the HDF galaxies which are our
preferred associations for the 15 ISO-HDF sources we take as having
reliable associations in the HDF region: these data
are used to compute the spectral energy distributions of the galaxies
in Paper V. 

The same procedure was then applied to an  $I_{814}$ band
HFF catalogue, resulting in the association of a further 11 ISO-HDF
sources with HFF objects (10 galaxies and 1 star): the construction
of this catalogue is described in the Appendix, where the HFF
associations are tabulated. These results, taken together, provide 
lower limits to the reliability of the four ISO-HDF samples; lower
limits because we have been conservative in accepting associations
as reliable, and a substantial number of further associations fall
just below our threshold, and might rise above it, for example, once
more is known of the field distortions in ISOCAM data. The 6.7$\mu$m
and 15$\mu$m complete samples are at least 71 per cent and 68 per cent 
reliable, respectively, while the reliabilities of the supplementary
6.7$\mu$m and 15$\mu$m samples are no worse than 35 per cent and
67 per cent respectively.

\subsection{Notes on individual associations}

In this subsection we discuss in detail the associations presented
in Table 1:
these comments should be borne in mind when using the results of
Table 1. In what follows, galaxies in the $I_{814}$ HDF
image are denoted by the names assigned
by Williams et al. (1996), prefixed by ``ST'', while those from the $H+K$ image of Cowie
et al. (1997) are denoted by their number in the catalogue of Songaila
(1997), with the prefix ``IfA'', and photometric redshifts have been
computed using the method of Mobasher et al. (1996), without using the
ISO data themselves.


	\begin{enumerate}

	\item {\bf ISOHDF3 J123641.1+621129}: The position of this 15$\mu$m 
	source falls just inside the Hubble Flanking Fields, but it is
	included here because its Airy disk encloses the very bright
	galaxy ST4-976.0, with which we have associated it: its
	proximity to the edge of the HDF means that ST4-976.0 has no
	counterpart in the Songaila (1997) catalogue, and we have
	estimated a photometric redshift of $z=0.047$ for it.

	\item {\bf ISOHDF2 J123641.6+621142}: This is associated with
	the brighter member (ST4-948.0) of a merging pair of galaxies,
	which is too near the edge of the HDF to be included in the
	near-infrared catalogue of Songaila (1997), despite appearing
	bright in both the $J$ and $H+K$ images of Cowie et
	al. (1997). Phillips et al. (1997) report a spectroscopic
	redshift of $z=0.585$ for ST4-948.0.

	\item {\bf ISOHDF2 J123642.6+621210}: This source falls midway
	between two spiral galaxies: ST4-656.0/IfA11 (\mbox{$z=0.454$},
	Songaila 1997) and ST4-795.0/IfA14 (\mbox{$z=0.432$}, Songaila 1997).
	It is associated with the former, which is brighter and yields
	a higher likelihood ratio, but note that
	this is one of the least reliable of our accepted
	associations.

	\item {\bf ISOHDF2 J123642.9+621309}: This interacting pair 
	(ST1-57.0 is the brighter member) falls in the Planetary
	Camera HDF field, and so is not included in the Songaila
	(1997) catalogue. We estimate a photometric redshift of
	$z=0.737$ for ST1-57.0.

	\item {\bf ISOHDF2 J123643.0+621152}: This source may have
	flux contributed by ST4-727.0/IfA59, as well as
	ST4-775.0/IfA20, as listed in Table 1. ST4-727.0/IfA59 is
	half a magnitude fainter in $I_{814}$ than ST4-775.0/IfA20,
	but is closer to the ISO position, and so yields a lower
	value of $P_{\rm ran}(I_{814})$: $P_{\rm ran}(I_{814})=0.269$ 
	as against $P_{\rm ran}(I_{814})=0.407$ for ST4-775.0/IfA20.
	Despite that we conservatively take ST4-775.0/IfA20 as being the
	association, on the basis of its lower photometric redshift:
	$z=0.820$, versus $z=1.63$ for ST4-727.0/IfA59. Having two
	galaxies with such low $P_{\rm ran}(I_{814})$ values clearly
	breaks the assumption, implicit in the likelihood ratio
	method, that there is not more than one true optical counterpart to 
	each ISO source, which gives some justification for
	over-riding our reliability criterion that  $P_{\rm
	ran}(I_{814})<0.35$ in this one case.

	\item {\bf ISOHDF3 J123643.7+621255}: This is associated with
	the brighter member (ST4-402.0/IfA32) of an interacting pair
	of galaxies on the basis of extremely reliable $I_{814}$
	data: $P_{\rm ran}(I_{814})=0.025$. 
	Cohen et al. (1996) give a spectroscopic
	redshift of $z=0.558$ for this galaxy.

	\item {\bf ISOHDF2 J123643.9+621130}: ST4-752.0 has an acceptable
	$P_{\rm ran}(I_{814})$ value. It is too close to the edge of the HDF to have been 
	included in the Songaila (1997) catalogue, although it looks
	bright in both the $J$ and $H+K$ images of Cowie et
	al. (1997). A spectroscopic redshift of $z=1.013$ is given for
	ST4-752.0 by Cohen et al. (1996).

	\item {\bf ISOHDF2 J123646.4+621406}: This source is
	confidently associated with ST2-251.0/IfA9: there are no
	plausible alternative associations. Cohen et al. (1996)
	quote a spectroscopic redshift of $z=0.960$ for this galaxy,
	and the broad emission lines in the spectrum shown by
	Songaila (1997) indicates that this galaxy hosts an AGN.

	\item {\bf ISOHDF2 J123647.1+621426}: This is identified as a
	stellar object (ST2-381.0). The $V_{606}$, $I_{814}$, $J$ and
	$H+K$ magnitudes can be fit very well with a $T=3450$ K
	blackbody, so this appears to be an M0 star. The corresponding
	predicted flux at 6.7$\mu$m would be 15$\mu$Jy, a factor of
	two lower than we observe, so we must presume that the star
	has a circumstellar dust shell, perhaps analogous to U Aur 
	(Rowan-Robinson \& Harris 1982).

	\item {\bf ISOHDF3 J123648.1+621432}: This source is just
	outside the HDF, but is included here because we have confidently
	associated it with a bright HDF galaxy (ST2-537.0): this
	galaxy is not included in the Songaila (1997) catalogue,
	because it is at the edge of the HDF. We estimate a
	photometric redshift of $z=0.023$ for this galaxy, using the 
	methods of Mobasher et al. (1996).

	\item {\bf ISOHDF2 J123648.2+621427}: ST2-537.0 is clipped off
	the edge of the Cowie et al. (1997) $H+K$ image: the
	(photometric) redshift for this galaxy is $z=0.023$, as given
	above.

	\item {\bf ISOHDF2 J123648.4+621215}: The flux from this
	source may be a combination of that from ST4-186.0/IfA44, an
	elliptical galaxy, as well as our preferred choice of
	ST4-260.111/IfA45, which
	is a spiral galaxy with a bright giant H{\sc ii} region. We favour
	the latter on the conservative basis of its having a lower
	photometric redshift ($z=0.778$ versus $z=1.512$), but note
	that both galaxies may contribute flux to this source.

	\item {\bf ISOHDF2 J123649.7+621315}: This source falls within
	a small group
	of bright galaxies, which are probably not physically
	associated. ST2-264.1/IfA17 is the preferred association for
	ISOHDF2 J123649.7+621315, although it is
	possible that this source also includes flux from 
	ST2-256.0/IfA43 and ST2-239.0/IfA35.  ST2-264.1/IfA17 has
	a spectroscopic redshift of $z=0.475$ (Cohen et al. 1996).

	\item{\bf ISOHDF3 J123649.8+621319}: This source lies in the
	same group of galaxies as ISOHDF2 J123649.7+621315 
	and, as with that source, we favour  ST2-264.1/IfA17 
	as the most likely association, but note that  
	it is likely that this 15$\mu$m source includes flux from
	ST2-256.0/IfA43 and ST2-239.0/IfA35, as well as, perhaps, 
	ST2-404.0/IfA6: as noted above, Cohen et al. (1996) quote
	a spectroscopic redshift of $z=0.475$ for ST2-264.1/IfA17.

	\item {\bf ISOHDF3 J123651.5+621357}: This is associated with 
	ST2-652.0/IfA10: Cohen et al. (1996) give the
	spectroscopic redshift of ST2-652.0/IfA10 as $z=0.557$.

	\item {\bf ISOHDF2 J123658.9+621248}: This source is
	associated with ST3-534.0/IfA27: there are no plausible
	alternative associations, and Cohen et al. (1996) have 
	determined its redshift spectroscopically to be $z=0.320$.
	
	\end{enumerate}


In Fig.3 we show the immediate surroundings (in an $I_{814}$ band mosaic) of 
the 15 ISO-HDF sources reliably associated with HDF galaxies. The dashed circle
in the centre of each plot marks the Airy disk of the ISO source,
and the square is centred on the position in the optical catalogue
of Williams et al. (1996) of the galaxy with which we have
associated it: for those plots where the edge of the HDF is within the
frame, the ISO-HDF source position is no longer placed at the
centre of the plot, while the presence of a  second circle 
in the same field indicates the
Airy disk of another ISO-HDF source from the same sample. Three-colour
($B_{450},V_{606},I_{814}$) versions of these plots can be viewed at 
{\tt http://artemis.ph.ic.ac.uk/hdf/catalogue.html}.

\subsection{Sources not reliably associated}

Twenty two ISO-HDF sources have not been reliably
associated with stars or galaxies in the optical HDF catalogue
of Williams et al. (1996) or in our own HFF catalogue: in (a) the 
complete 6.7$\mu$m sample: 
123655.1+621423 and 123658.8+621313; (b)  the supplementary 6.7$\mu$m
sample: 123641.5+621309,  123642.5+621256, 123643.1+621203,
123646.4+621440, 123648.6+621123, 123650.2+621139, 123655.2+621413, 
123655.7+621427, 123656.1+621303, 123656.6+621307, 123657.6+621205, 
123658.6+621309 and 123701.2+621307; (c) the complete 15$\mu$m sample:
123634.3+621238, 123637.5+621109, 123646.9+621045, 123653.6+621140, 
123659.4+621337 and 123702.5+621406; and (d) the  supplementary
15$\mu$m sample: 123658.1+621458.

A number of these sources have likeliest associations 
that lie on the sharply falling portion of the curve of
$P_{\rm ran}(I_{814})$ against $LR$, and may possibly rise above
our reliability threshold once a more accurate model for $e(x,y)$
in equations (1) and (2) can be computed, properly taking into 
account the as yet uncertain field distortion in ISOCAM data and
improving the astrometric accuracy of the ISO-HDF maps.

\section{DISCUSSION AND CONCLUSIONS}

We have conservatively associated fifteen ISO-HDF sources detected
at 6.7$\mu$m or 15$\mu$m with optical galaxies in the HDF
catalogue of Williams et al. (1996), eight of which are also in the
near-infrared catalogue of Songaila (1997): a further association is 
made with a star. This was done using two
independent procedures, namely the likelihood ratio method
(Sutherland \& Saunders 1992) and visual inspection. These gave
consistent results, whose reliability we tested by computing
the likelihood ratios for galaxies to be associated with
fictitious sources placed at random in the Hubble Deep Field. A
similar procedure yielded a further eleven associations with
objects (ten galaxies and one star) in the Hubble Flanking
Fields: more details of this are given in the Appendix.

We detect 10 of the 44 brightest $I_{814}$ band objects in the Williams et 
al. (1996) catalogue (i.e. those with $I_{814}<22.04$): 8 of these 44
objects are stars, which we discuss no further. Of the 36 galaxies,
we detect 13 per cent (2 out of 15) of the ellipticals, 30 per cent (6/18) of
the spirals and 67 per cent (2/3) of the irregulars/mergers.
We divide these 36 galaxies into three bins of 12 galaxies each for
redshift and the three optical colours $V_{606}-I_{814}$,
$B_{450}-V_{606}$, and $U_{300}-B_{450}$. There
are (3,4,3) of our galaxies in bins of increasing redshift, so the
galaxies associated with the ISO-HDF sources have a similar redshift
distribution to bright HDF galaxies in general: 5
out if 10 have redshifts greater than 0.5.
We find (4,4,2) of our objects in the bins of increasing
$V_{606}-I_{814}$ and of increasing $U_{300}-B_{450}$, and (4,4,3) in
bins of increasing $B_{450}-V_{606}$.
A detailed study of the properties of the galaxies associated
with ISO-HDF source, contrasting them with those of the HDF
galaxy population as a whole, will be the topic of a later
paper in this series, but is clear, though, that, amongst bright HDF
galaxies,  ISO has a tendency to detect luminous, star-forming galaxies
at fairly high redshift and with disturbed morphologies, in preference
to nearby ellipticals: the
implications of this result is discussed in Paper V.

Further information on the ISO-HDF project can be found on the ISO-HDF
WWW pages: see {\tt http://artemis.ph.ic.ac.uk/hdf/}.

\section*{ACKNOWLEDGMENTS}
This paper is based on observations with ISO, an ESA project, with
instruments funded by ESA Member States (especially the PI countries:
France, Germany, the Netherlands and the United Kingdom) and with
participation of ISAS and NASA.
This work was supported by PPARC grant GR/K98728 and by the EC TMR
Network FMRX-CT96-0068. We thank the referee, Harry Ferguson, for
many helpful comments.

\section*{REFERENCES}

\myref{Cesarsky C., et al., 1996, A\&A, 315, 32}
\myref{Cohen J.G et al., 1996, ApJ, 471, L5}
\myref{Cowie  L. L.,  Clowe D., Fulton E., Cohen J.G., Hu E.M., Songaila A., Hogg D.W., Hodapp K.W.,
1997, in preparation}  
\myref{Draper P.W., Eaton N., 1996, {\sc pisa}, Starlink User Note
109, {\tt http://star-www.rl.ac.uk/star/docs/sun109.htx 
        /sun109.html}}
\myref{Gallego J., Guzman R., 1997, {\tt
http://www.ucolick.org/}}$\;\tilde{}${\tt deep/hdf/hdf.html}
\myref{Goldschmidt P., et al., 1997, MNRAS, in press (Paper II)}
\myref{Fomalont E.B., Kellermann K.I., Richards E., Windhorst R.A.,
Partridge R.B., 1997, ApJ, 475, 5}
\myref{Kessler M., et al., 1996, A\&A, 315, 27}
\myref{Lawrence A., Rowan-Robinson M., Leech K.J., Jones D.H.P, Wall
J.V., 1989, MNRAS, 240, 329}
\myref{Lowenthal J.D. et al., 1997, in preparation}
\myref{Mobasher B., Rowan-Robinson M., Georgakakis A., Eaton N., 1996,
MNRAS, 282, L7}
\myref{Moustakas L., Zepf S., Davis M., 1997, 
{\tt http://astro.berkeley.edu/davisgrp/HDF/observations.html}}
\myref{Oliver S.J., et al., 1997, MNRAS, in press (Papper III)}
\myref{Phillips A.C., Guzman R., Gallego J., Koo D.C., Lowenthal J.D.,
	Vogt N.P., Faber S.M., Illingworth G.D., 1997, ApJ, submitted}
\myref{Rowan-Robinson M., Harris S., 1982, MNRAS, 200, 197}
\myref{Rowan-Robinson M., et al., 1997, MNRAS, in press (Paper V)}
\myref{Serjeant, S., et al., 1997, MNRAS, in press (Paper I)}
\myref{Songaila, A., 1997, ` The Hawaii Active Catalog of the Hubble 
Deep Field'
{\tt http://www.ifa.hawaii.edu/}$\;\tilde{}${\tt cowie/tts/tts.html (15
February 1997)}}
\myref{Sutherland, W., Saunders W., 1992, MNRAS, 259, 413}
\myref{Williams, R.E., et al., 1996, AJ, 112, 1335}

\appendix
\setcounter{table}{0}
\renewcommand\thesection{\Alph{section}}
\renewcommand\thetable{\thesection\arabic{table}}

\section{FLANKING FIELD ASSOCIATIONS}	

\begin{table*}

\begin{minipage}{150mm}

\caption{Properties of the galaxies reliably associated with the
ISO-HDF sources in the Hubble Flanking Fields}

\begin{tabular}{lcccccccc}
ISO Source & RA & Dec & $P_{\rm ran}(I_{814})$ & $I_{814}$ & 6.7$\mu$m & 15$\mu$m & 8.4 GHz & z \\

\hline

J123633.9+621217 3C& 12 36 34.4 & +62 12 13.9 & 0.032 & 18.98 & - & 726 &40.0  &- \\ 
J123635.9+621134 3C& 12 36 36.8 & +62 11 35.5 & 0.005 & 17.97 & - & 420 &- &0.078 \\
J123636.5+621348 3C& 12 36 36.9 & +62 13 46.2 & 0.243 & 21.63 & - & 649 &- &- \\
J123639.3+621250 3C& 12 36 40.0 & +62 12 50.2 & 0.192 & 21.18 & $<97.57$& 433 &- & \\
J123653.0+621116 3C& 12 36 53.2 & +62 11 17.5 & 0.203 & 21.47 & - & 327 &- &- \\
J123657.4+621414 2S& 12 36 57.7 & +62 14 18.6 & 0.272 & 22.25 &38.1 &$<243$ &- &- \\
J123658.7+621212 3C& 12 36 59.0 & +62 12 09.1 & 0.267 & 21.76 &$<89.1$& 336 &- &- \\
J123700.2+621455 3C& 12 36 59.8 & +62 14 50.5 & 0.282 & 21.52 & - & 291 &- &- \\
J123702.0+621127 3S& 12 37 02.0 & +62 11 23.0 & 0.032 & 18.92 & - & 326 &- &0.136 \\
J123705.7+621157 3C& 12 37 05.9 & +62 11 53.8 & 0.161 & 20.96 & - & 472 &- &0.904 \\

\hline

\end{tabular}

\medskip
The first column gives the ISO source name, as listed in Paper II, 
minus ``ISOHDF'' and the prefix indicating in which band the source 
was detected. This latter information is indicated by ``2'' (LW2
filter, 6.7$\mu$m) and ``3'' (LW3 filter, 15$\mu$m) following the
source name: ``C'' or ``S'' indicate whether the ISO source comes
from the complete or supplementary sample for that waveband.
The next columns give the RA and Dec of the associated galaxy,
in J2000 coordinates, followed by the value of $P_{\rm ran}(I_{814})$
for the association, and the $I_{814}$ total magnitude of the
associated object. Following that are the source fluxes in the two
ISO bands (in $\mu$mJy), after which is the 8.4 Ghz radio flux from
Fomalont et al. (1997), also in $\mu$mJy. The final column lists the 
spectroscopic
redshifts of the three galaxies for which they have been measured:
those for ISOHDF3C J123702.0+621127 and ISOHDF3C J123705.7+621157 come
from Phillips et al. (1997), while that for ISOHDF3C J123635.9+621134
is from Moustakas, Zepf \& Davis (1997). One reliable association is
omitted from this Table: ISOHDF3C J123709.8+621239 is associated
with a star.

\end{minipage}

\end{table*}

In this Appendix we discuss the association of ISO-HDF sources in the 
Hubble Flanking Fields (HFF) region. We constructed a $I_{814}$ band
catalogue of the HFF using the connected pixel algorithm {\sc pisa}
(Draper \& Eaton 1996). No attempt was made to push the detection
threshold to include the faintest objects in the HFF images, as 
we knew that we were principally interested in the brighter galaxies
as possible associations for ISO  sources, on the basis of 
associations made in the HDF. We used {\sc pisa} to compute total $AB$
magnitudes for the detected objects, which it does using a
curve-of-growth estimator. The limiting magnitude of detected objects
varied from field to field, and was conservatively set at
$I_{814}$=24, which corresponds to the depth of the shallowest field:
objects fainter than that were neglected in the association procedure.

The photometric calibration of our HFF catalogue was checked by 
comparing magnitudes of galaxies in the small overlap between the
HDF and HFF, and by comparing magnitudes for those flanking field
galaxies for which the DEEP collaboration (Gallego \& Guzman 1997)
have measured redshifts. In neither case was there evidence of a
systematic offset, apart from a trend for the DEEP magnitudes
to be brighter than our {\sc PISA} magnitudes at the magnitude
limit of our catalogue by $\sim0.2$ mag: this has no bearing on the
likelihood ratio analysis.

Associations were made with objects in this catalogue using the same
likelihood ratio method as described in Section 3. A total of eleven
ISO-HDF sources were reliably associated (using the same reliability 
criterion as before, i.e. $P_{\rm ran}(I_{814})<0.35$) with objects in
the flanking field catalogue. These are tabulated in Table A1.

\end{document}